\documentclass[letterpaper, 10 pt, conference]{ieeeconf}
\IEEEoverridecommandlockouts                              % This command is only needed if
\usepackage{latexsym,amssymb, color}
\usepackage{epsfig}
\usepackage{mathrsfs}
\usepackage{amsmath}
\usepackage{amssymb}
\usepackage{indentfirst,latexsym,bm}
\usepackage{subfigure}
\usepackage{graphicx}
\usepackage{graphics}
\usepackage{CJK}

\newtheorem{prop}{Proposition}
\newtheorem{exx}{Example}
\newtheorem{remm}{Remark}
\newenvironment{remark}{\begin{remm}\rm }{\hfill \hspace*{1pt} \hfill $\lrcorner$\end{remm}}
\newenvironment{proposition}{\begin{prop} \rm }{\hfill \hspace*{1pt} \hfill $\lrcorner$\end{prop}}
\newenvironment{example}{\begin{exx}\rm }{\hfill \hspace*{1pt} \hfill $\lrcorner$ \end{exx}}
\newcommand{\col}{\mbox{col} }

\def\L2{{\cal L}_2}
\def\L2e{{\cal L}_{2e}}

\def\rea{\mathbb{R}}

\def\begequarr{\begin{eqnarray}}
\def\endequarr{\end{eqnarray}}
\def\begequarrs{\begin{eqnarray*}}
\def\endequarrs{\end{eqnarray*}}
\def\begarr{\begin{array}}
\def\endarr{\end{array}}
\def\begequ{\begin{equation}}
\def\endequ{\end{equation}}
\def\lab{\label}
\def\begdes{\begin{description}}
\def\enddes{\end{description}}
\def\begenu{\begin{enumerate}}
\def\begite{\begin{itemize}}
\def\endite{\end{itemize}}
\def\endenu{\end{enumerate}}

\def\lef[{\left[\begin{array}}
\def\rig]{\end{array}\right]}

\def\begcen{\begin{center}}
\def\endcen{\end{center}}
\def\begrem{\begin{remark}\rm}
\def\endrem{\end{remark}}
\def\ba{\begin{array}}
\def\ea{\end{array}}

\def\calM{{\cal M}}

\def\calM{{\cal M}}

\def\rea{\mathbb{R}}

\begin{document}
%%%%%%%%%%%%%%%%%%%%%%%%%%%%%%%%%%%%%%%%%%%%%%%%%%%%%%%%%%%%%%
\title{\LARGE \bf Immersion and Invariance Stabilization of Nonlinear Systems: A Horizontal Contraction Approach }

%\author{Lei Wang, Fulvio Forni, Romeo Ortega, Hongye Su}
\date{}
\author{Lei Wang$^{1}$, Fulvio Forni$^2$, Romeo Ortega$^{3}$, and Hongye Su${^4}$% <-this % stops a space
\thanks{*This work is partially supported by National Basic Research Program of China (973 Program 2013CB035406) ; National Natural
Science Foundation of China (NSFC: 61134007 and 61320106009). (\emph{Corresponding author: Hongye Su.})}% <-this % stops a space
\thanks{$^{1}$Lei Wang is with State Key Laboratory of Industrial Control Technology, Institute of Cyber-Systems and Control, Zhejiang University, 310027, Hangzhou, P.R.China and CASY - DEI, University of Bologna, Italy
        {\tt\small lwang@iipc.zju.edu.cn}}%
\thanks{$^{2}$F. Forni is with the University of Cambridge, Department of Engineering,
Trumpington Street, Cambridge CB2 1PZ, and with the Department
of Electrical Engineering and Computer Science,
University of Li{\`e}ge, 4000 Li{\`e}ge, Belgium,
        {\tt\small ff286@cam.ac.uk}.
Research supported by FNRS.
The paper presents research results of the Belgian Network DYSCO
(Dynamical Systems, Control, and Optimization), funded by the
Interuniversity Attraction Poles Programme, initiated by the Belgian
State, Science Policy Office. The scientific responsibility rests with
its authors.}%
\thanks{$^{3}$Romeo Ortega is with Laboratoire des Signaux et Syst$\grave{e}$mes, CNRS UMR 8506, Sup$\acute{e}$lec, 91192 Gif sur Yvette, France
        {\tt\small ortega@lss.supelec.fr}}%
\thanks{$^{4}$Hongye Su is with State Key Laboratory of Industrial Control Technology, Institute of Cyber-Systems and Control, Zhejiang University, 310027, Hangzhou, P.R.China
        {\tt\small hysu@iipc.zju.edu.cn}}%
}
\maketitle
%%%%%%%%%%%%%%%%%%%%%%%%%%%%%%%%%%%%%%
\begin{abstract}
The main objective of this paper is to propose an alternative procedure to carry out one of the key steps  of immersion and invariance stabilising controller design. Namely, the one that ensures attractivity of the manifold whose internal dynamics contains a copy of the desired system behaviour. Towards this end we invoke  {\em contraction theory} principles and ensure the attractivity of the manifold rendering it {\em horizontally contractive}. The main advantage of adopting this alternative approach is to make more systematic the last step of the design with more explicit degrees of freedom to accomplish the task. The classical case of systems in feedback form is used to illustrate the proposed controller design.
\end{abstract}


%%%%%%%%%%%%%%%%%%%%%%%%%%%%%%%%%%%%%%%%%%%%%%%%%%%%%%%%%%

\section{Introduction}
Immersion and invariance (I$\&$I) is a controller design technique that has been recently proposed in the literature to stabilise non-linear systems \cite{astort}---see also the recent book \cite{astkarort} where many applications of the technique are presented.
{The I$\&$I approach captures
the desired behaviour of the system to be controlled by introducing a target dynamical system.
Then, a suitable stabilizing control law is designed to
guarantee that the controlled system asymptotically behaves like the target system}.
More precisely, the I$\&$I methodology relies on finding
a manifold in the plants state--space that can be rendered {\em invariant and attractive} {by feedback control},
{such that  (i)~on the manifold, the closed loop dynamics behaves like the desired dynamics
(ii)~away from the manifold, the control law steers the state of the system towards the manifold.
The usual way to}
carry out the latter step {is to} define an extended dynamical system {given by a copy of the plant
and by a new \emph{error} dynamics, denoted by the coordinate $z$,} that measures the distance to the manifold.
{Then, a} full--state feedback controller must be designed to ensure boundedness of the plant state and convergence to zero of the {$z$} coordinate.
{The main stabilisation result in I$\&$I states that the evaluation of this control law on the manifold
defines an asymptotically stabilising controller for the system. The construction leads to a static controller,
since the control law is a function only of the plant state.}

The design of the aforementioned full--state feedback controller is {not} systematic
{and finding a controller that renders the desired manifold attractive
could be challenging in practice.}
The main objective of this paper is to carry out this step by {exploiting} {\em contraction theory} principles \cite{sloloh}.
More precisely, {we will use \emph{horizontal contraction} \cite{forsep} to draw geometric conditions
that guarantee the attractiveness of the desired manifold.}
The main advantage of adopting this alternative approach in I$\&$I is to make more systematic
{the design of the control action away from the desired manifold.
We anticipate that the stabilization of the extended system of I$\&$I is replaced by the
stabilization of the prolonged system \cite{Crouch1987}, defined by the plants and its linearization.
In comparison to I$\&$I, the local nature of the approach pursued in this paper
provides more degrees of freedom in the design of the controller,
possibly widening the use of I$\&$I in applications.}

The paper is organized as follows. Section \ref{sec2} briefly recalls the standard I$\&$I controller design procedure.
{The novel design based on horizontal contraction is illustrated in Section \ref{sec3}. The section provides the main result
of the paper, whose proof is in Appendix B}.  The classical example of systems in feedback form is  presented in Section \ref{sec4}.
Concluding remarks are detailed in Section \ref{sec5}. Appendix A contains a counterexample to the classical I$\&$I design.
{This minor issue of the classical I$\&$I is easily fixed by enforcing a simple extra assumption.}

\noindent {\bf Notation} For $x \in \rea^n$ we denote the Euclidean norm $|x|^2:=x^\top x$.  Given a function $f:  \rea^n \to \rea$ we define the differential operators
$$
\nabla f:=\left(\frac{\displaystyle \partial f }{\displaystyle \partial x}\right)^\top,\;\nabla_{x_i} f:=\left(\frac{\displaystyle
\partial f }{\displaystyle \partial x_i}\right)^\top,
$$
where $x_i \in \rea^p$ is an element of the vector $x$. For a mapping $g : \rea^n \to \rea^m$, its Jacobian matrix is defined as
$$\nabla g:=\left [\begin{array}{cc}(\nabla g_1)^\top \\
\vdots\\ (\nabla g_m)^\top \end{array}\right],$$ where $g_i:\rea^n \to \rea$ is the $i$-th element of $g$.
When clear from the context the subindex of the operator $\nabla$ and the arguments of the functions will be omitted.  All the functions in the paper are assumed sufficiently smooth.

%
%%%%%%%%%%%%%%%%%%%%
\section{The Standard I$\&$I Stabilisation Procedure}
\lab{sec2}
%%%%%%%%%%%%%%%
%
Consider the system
\begequ
\label{sys}
\dot x = f(x) + g(x)u
\endequ
with state $x\in\mathbb{R}^n$, the input control $u\in\mathbb{R}^m$, and an assignable equilibrium point
$$
x_*\in\{x \in \mathbb{R}^n\;|\;g^\perp(x)f(x)=0\}
$$
to be {\em stabilized}, where $g^\perp: \rea^n \to \rea^{(n-m) \times m}$ is a full--rank left annihilator of $g(x)$. Stabilisation is achieved in I$\&$I
fulfilling the {following four steps. The reader is referred to \cite{astkarort} for the proof of the next proposition}

\begin{proposition}
\label{pro1}
Assume that there exist mappings
$$
\ba{rcl}
\alpha: \mathbb{R}^p\rightarrow\mathbb{R}^p\,,\quad \pi: \mathbb{R}^p\rightarrow\mathbb{R}^n\,,\quad c: \mathbb{R}^n\rightarrow\mathbb{R}^m,\;\\
\phi: \mathbb{R}^n\rightarrow\mathbb{R}^{n-p}\,,\quad v: \mathbb{R}^n\times\mathbb{R}^{n-p}\rightarrow\mathbb{R}^{m}\,,
\ea
$$
with $p<n$, such that the following hold.
\begin{itemize}
\item[(A1)]
\emph{(Target system)}
The  system
\begequ
\label{tsys}
	\dot \xi = \alpha(\xi) \ ,
\endequ
has a globally asymptotically stable equilibrium at $\xi_* \in \rea^p$ and $x_* = \pi(\xi_*)$.

\item[(A2)]
\emph{(Manifold invariance condition)}
For all $\xi\in\mathbb{R}^p$,
\begequ
\label{fbi}
	f(\pi(\xi))+g(\pi(\xi))c(\pi(\xi))=\nabla \pi(\xi)\alpha(\xi) \ .
\endequ

\item[(A3)]
\emph{(Implicit manifold description)}
The following set identity holds
\begequ
\label{manifold}
	\mathcal{M}:=\{x\in\mathbb{R}^n | x=\pi(\xi)\} = \{x\in\mathbb{R}^n | \phi(x)=0\} \ .
\endequ
 \item[(A4)]
 \emph{(Manifold attractivity and trajectory boundedness)}
 Consider the system
 \begequarr
 \lab{xdyn}
 \dot x & = & f(x)+g(x)v(x,z)\\
 \lab{zdyn}
 \dot z & = & \nabla \phi(x)[ f(x)+g(x)v(x,z)],
 \endequarr
 with the initial condition constraint
 \begequ
\lab{zzer}
z(0)=\phi(x(0)),
\endequ
and $v(x,z)$ verifying
\begequ
\lab{v}
v(\pi(\xi),0)=c(\pi(\xi)),\;\forall \xi \in \rea^p.
\endequ
All trajectories of the system are bounded and satisfy
 \begequ
 \lim_{t\to\infty}z(t)=0\,.
 \endequ
 \end{itemize}
Then, $x_*$ is a  {\em globally asymptotically stable} (GAS) equilibrium of the closed--loop system
\begequ
\label{sys-cl}
\dot x = f(x) + g(x)v(x,\phi(x))\,.
\endequ
\end{proposition}

{Following the discussion in the introduction,
the accomplishment of step (A4) is not systematic and may challenge the successful completion of the I$\&$I design.
Exploiting \cite{forsep},  we propose in the next section to replace (A4) by  a novel condition
based on horizontal contraction  \cite{forsep}.}

\begrem
In comparison to
the results presented in \cite{astort,astkarort}, we have added the initial condition constraint \eqref{zzer} and the requirement \eqref{v}. The first condition ensures that $z(t)=\phi(x(t)),\;\forall t \geq 0$, while the second one guarantees that the $x$--system behaves like the $\xi$--system when restricted to the manifold $\mathcal{M}$. These requirements were implicitly assumed in previous works. If these conditions are not imposed it is possible to show that the claim of Proposition \ref{pro1} is false.  An example of this fact is given in Appendix A.
\endrem

%{
%\begrem
%In \cite{Seibert}, reduction principles for (asymptotic) stability is proposed by means of rendering a compact, positively invariant set, whose idea is substantially different from the I$\&$I method. Note that the invariant manifold in I$\&$I technique is generalized by the target system.
%\endrem
%}
{
\begrem
The I$\&$I technique makes contact with the literature of
invariant manifolds stability \cite{Wiggins1994} and of
conditional stability (relative to a set) \cite{Seibert}.
Indeed, in the I$\&$I technique the action of a state-feedback
controller renders invariant and stabilizes a
suitable submanifold of the system state space
while enforcing a desired steady-state behavior, represented
by the target dynamics.
\endrem
}

%
%%%%%%%%%
\section{The I$\&$I Horizontal Contraction Procedure}
\lab{sec3}
%%%%%%%%
%
The proposition below proposes to replace the step (A4) in Proposition \ref{pro1} by a horizontal contraction based design that ensures attractivity of the manifold $\mathcal{M}$.
\begin{proposition}
\lab{pro2}
Given the conditions (A1)--(A3) in Proposition \ref{pro1}, assume there exist mappings
{$$
\begin{array}{ll}
P :\rea^n \to  \rea^{({n-p}) \times ({n-p})}, & \ P=P^T > 0 \\
R: \rea^n \to \rea^{(n-p) \times n}, & \\
\;\beta:\rea^n \to \rea^m , & \\
\rho:\rea^n \to \rea , & \ \rho > 0 \ ,
\end{array}
$$
}
such that the following holds.\\
{(A4') \emph{(Manifold attractivity via horizontal contraction)}
}
\begenu
\item[(i)] For all $x \in \rea^n$, $R(x)$ is full rank and
\begin{equation}
\label{eq:cond_R}
R(\pi(\xi)) = \nabla \phi(\pi(\xi)),\;\forall \xi \in \rea^p.
\end{equation}
\item[(ii)] For all $\xi \in \rea^p$
$$
\beta(\pi(\xi))=c(\pi(\xi)).
$$
\item[(iii)] The candidate Finsler-Lyapunov function
$V:\mathbb{R}^n\times\mathbb{R}^n \to \rea_{\geq 0}$  given by
\begequ
\lab{finlya}
V(x,\delta x) := \delta x^\top R^\top(x) P(x) R(x) \delta x,
\endequ
satisfies
\begin{equation}
\label{eq:dotV}
\dot{V}(x,\delta x) \leq -\rho(x) V(x,\delta x)
\end{equation}
along the trajectories of the prolonged system
\begequarr
\lab{cloloo}
\dot{x} & = & f(x) + g(x) \beta(x) \\
\lab{varsys}
\dot{\delta x} & = & \nabla[f(x)+g(x)\beta(x)] \delta x.
\endequarr

\item[(iv)] The trajectories of  \eqref{cloloo} are {\em bounded}.
\endenu

Then, $x_*$ is a GAS equilibrium point of the closed--loop system  \eqref{cloloo}.
Furthermore, if the fixed point $\xi_*$ of the target system \eqref{tsys} is
{hyperbolic}, then $x_*$ is hyperbolic.
\end{proposition}

{
\begin{remark}
A natural simple choice for $R(x)$ is $\nabla \phi(x)$, provided that $\nabla \phi(x)$ is full rank for all $x\in\mathbb{R}^n$.
%{Also, in some cases, the controller $c(\pi(\xi))$ that renders the manifold invariant, can ensure the horizontal
%contraction condition (iii)---then, setting $\beta(x)=c(x)$ obviates the need to verify condition (ii).}
\end{remark}
}

\begin{remark}
In contrast with the classical I{\&}I Proposition \ref{pro2} directly provides the static state--feedback controller $\beta(x)$.
{This should be compared with the control $v(x,z)$ that should verify condition (A4) for the augmented system  \eqref{xdyn}, \eqref{zdyn}, which is later evaluated on the manifold to generate the actual control to be applied, that is, $v(x,\phi(x))$.}
\end{remark}

\begin{remark}
Proposition \ref{pro2} can be formulated in a similar way for any forward invariant region $\mathcal{C} \subseteq \rea^n$. If $\mathcal{C} = \rea^n$ then, as stated in the proposition, one gets GAS. Otherwise, one gets regional stability. This formulation may be useful in applications when global results are difficult to achieve or when the system lives in a manifold different from $\rea^n$. Note that if $\mathcal{C}$ is compact, then the condition (iv) of boundedness of trajectories is automatically satisfied.
\end{remark}

%
%%%%%%%%
\section{Application to Systems in Feedback Form}
\lab{sec4}
%%%%%%%%%%%%
%
Consider the class of systems in feedback form described by the equations
\begequ
\ba{rcl}
\label{feedbackform}
\dot x_1 &=& f(x_1,x_2)\,,\\
\dot x_2 &=& u\,,
\ea
\endequ
with $x:=\col(x_1,x_2)\in\mathbb{R}^n\times\mathbb{R}$, and $u\in\mathbb{R}$. Consistent with the standard backstepping scenario \cite{kkk} assume there exists a mapping $\pi_2:\mathbb{R}^n\rightarrow\mathbb{R}$ such that the system
$$
\dot x_1=f(x_1,\pi_2(x_1))
$$
has a GAS equilibrium at the origin. A sensible choice of the target dynamics is then given by
\begequ
\label{tar-feedback}
\dot\xi = f(\xi,\pi_2(\xi))\,,
\endequ
and this implies that the mapping $\pi(\xi)$ has the form
$$
\pi(\xi) = \left[
             \begin{array}{c}
               \xi \\
               \pi_2(\xi) \\
             \end{array}
           \right]\,.
$$
To verify Assumptions (A2) and (A3) of Proposition \ref{pro1} we can choose
\begequarr
\lab{c}
c(\xi,\pi_2(\xi))&=&\nabla\pi_2(\xi)f(\xi,\pi_2(\xi))\\
\lab{phi}
\phi(x)&=&x_2-\pi_2(x_1),
\endequarr
which clearly satisfy (\ref{fbi}) and (\ref{manifold}).

The differential relation of the system (\ref{feedbackform}) in closed--loop with the control $\beta(x)$ is
$$
\dot{{\delta x}} = \left[
                               \begin{array}{cc}
                                 \nabla_{x_1}f(x) & \nabla_{x_2}f(x) \\
                                 \nabla_{x_1}\beta(x) & \nabla_{x_2}\beta(x) \\
                               \end{array}
                             \right]\delta x
                             =: Q(x) \delta x,
$$
 From Proposition \ref{pro2} we select $R(x)=\nabla \phi(x)$ and $P(x)=I$. Whence the Finsler--Lyapunov function \eqref{finlya} takes the form
\begequ
\label{V-feedback}
V(x,\delta x) = \delta x^{\top}M(x_1)\delta x \,,
\endequ
where we have defined
\begequ
\label{M-feedback}
\ba{rcl}
M(x_1)&:=&[\nabla\phi(x)]^{\top}\nabla\phi(x)\\
       &=&                     \left[
                                 \begin{array}{cc}
                                 \nabla\pi_2(x_1) [\nabla\pi_2(x_1)]^{\top} & -\nabla\pi_2(x_1) \\
                                   -[\nabla\pi_2(x_1)]^\top & 1 \\
                                 \end{array}
                               \right]
\ea
\endequ
Fixing $\rho(x)=k>0$ the condition \eqref{eq:dotV} is satisfied if and only if
\begequ
\label{dotV-feedback}
\dot M(x_1) + M(x_1)[Q(x)+{k \over 2} I] + [Q^{\top}(x)+{k \over 2} I] M(x_1) \leq 0.
\endequ
We are in position to state the following proposition.
%
%%%%%%%%
\begin{proposition}
\label{pro3}
Consider a system described by equations of the form (\ref{feedbackform}) and suppose there exist mappings $\pi_2:\rea^n \to \rea$ and  $\beta:\rea^{(n +1)}\to \rea$ such that the following holds.
\begenu
\item[(a)] The system
$$
\dot x_1=f(x_1,\pi_2(x_1))
$$
has a GAS equilibrium at zero.
\item[(b)]  The inequality \eqref{dotV-feedback} is satisfied for some $k>0$.
\item[(c)] For all $\xi \in \rea$
$$
\beta(\xi,\pi_2(\xi))=\nabla\pi_2(\xi)f(\xi,\pi_2(\xi)).
$$
\endenu
{Then,}
the system  (\ref{feedbackform}) in closed--loop with $\beta(x)$ has a GAS equilibrium at zero.
\end{proposition}
%%%%%%

\begin{example}
To illustrate the result in Proposition \ref{pro3}, consider the two-dimensional system
\begequ
\label{ex-feedback}
\ba{rcl}
\dot x_1 &=& -x_1+\lambda x_1^3x_2\\
\dot x_2 &=& u\,,
\ea
\endequ
in which $\lambda>0$. We proceed now to verify condition (a). Selecting $\pi_2(x_1)=-x_1^2$ we obtain
$$
\dot x_1=f(x_1,\pi_2(x_1))=-x_1-\lambda x_1^5
$$
which has a GAS equilibrium at zero. To check condition (b) we, first, compute
\begequarrs
\phi(x) & = & x_1^2+x_2\\
M(x) & = & \left[ \begin{array}{cc} 4x_1^2 & 2x_1 \\ 2x_1 & 1 \\\end{array} \right]\\
Q(x) & = & \left[
     \begin{array}{cc}
       -1+3\lambda x_1^2x_2 & \lambda x_1^3 \\
       \nabla_{x_1}\beta(x) & \nabla_{x_2}\beta(x) \\
     \end{array}
   \right].
\endequarrs
Some lengthy, but straightforward calculations, show that
\begequ
\lab{bet}
\beta(x) = -{1\over2}(k-4)x_1^2 - {1\over2}kx_2-2\lambda x_1^4x_2\,
\endequ
solves (\ref{dotV-feedback}) with identity. It only remains to verify condition (c), which holds true because
$$
\beta(\xi,\pi_2(\xi))=2\xi^2(1+\lambda \xi^4)=\underbrace{(-2\xi)}_{\nabla\pi_2(\xi)}\underbrace{[ -\xi+\lambda \xi^3 (-\xi^2)]}_{f(\xi,\pi_2(\xi))}.
$$
In conclusion, the system \eqref{ex-feedback} in closed--loop with the control \eqref{bet} has a GAS equilibrium at the origin.
\end{example}

%
%%%%%%%%%%%
\section{Conclusions}
\lab{sec5}
%%%%%%%%%
%
An alternative procedure to complete the design of I$\&$I controllers for stabilization of nonlinear systems has been proposed. The central idea is
to replace by a contraction--based design the {stabilization step on the extended dynamics \eqref{xdyn},\eqref{zdyn}
required by condition (A4) of the I{\&}I procedure.}
%non-standard and sometimes cumbersome task of ensuring boundedness of the plant state and convergence to zero of the off--the--manifold coordinate.
The  main advantage of the {contraction--based} approach is to render more systematic the design and to give more degrees of freedom for its accomplishment.
The key step of the novel design is the use of horizontal Finsler--Lyapunov functions \cite{forsep}
that decays along the trajectories of the prolonged system,
in the spirit of classical Lyapunov theory.  Of course, similarly to all constructive procedures for the design of nonlinear controllers or observers,
for the successful application of the novel design proposed by the paper
it is necessary to solve a partial differential equation.
In particular, {for systems in feedback form}, it is necessary to find a controller $\beta(x)$ that satisfies \eqref{dotV-feedback}
($\beta(x)$ is encoded in $Q(x)$) for a suitable choice of $R(x)$ and $P$.

From the conceptual viewpoint,
{the use of Finsler-Lyapunov functions replaces the stabilization of the off-manifold coordinate $z$
of I$\&$I with the horizontal stabilization of the linearization along trajectories.
For instance, the method proposed in this paper stabilizes the \emph{linearization} of the system
along suitable directions of its tangent space, thus
providing a local and intrinsic feedback design procedure that
does not require any a-priori definition of the off-manifold coordinate $z$.
The advantage is a more general design method, possibly.
This generality is directly encoded into the conditions of Proposition \ref{pro2}:
the $z$ coordinate of classical I$\&$I
is replaced at local level by the matrix $R(x)$, which is one of the free parameters to be selected in the formulation of
the partial differential equation \eqref{dotV-feedback} ($M(x)$ depends on $R(x)$).
The intrinsic nature of the design combined with the increased degrees of freedom make}
the present formulation of horizontal contraction--based I$\&$I a promising {stabilization tool for applications}.

%
%%%%%%%%

%
%%%%%%%%%%%
\appendix
\subsection{Counterexample to Theorem 2.1 of \cite{astkarort} }
\lab{appa}
%%%%%%%%%
%

Our objective in this appendix is to show that if we follow all the steps of the standard I$\&$I procedure of Proposition \ref{pro1}---{\em without} imposing the conditions \eqref{zzer} and \eqref{v}---we cannot guarantee GAS of the equilibrium. Towards this end, consider the two--dimensional, linear, time--invariant system
$$
\ba{rcl}
\dot x_1 &=& -x_1 + \theta + u_1\,\\
\dot x_2 &=& u_2\,
\ea
$$
where $ \theta \neq0$ is a constant parameter and $u=\col(u_1,\,u_2)$ is the control input. The control objective is to stabilize the system at the origin using the I$\&$I procedure.

First, we select  the target system as $\dot\xi=-\xi$, which clearly has a GAS equilibrium at zero, verifying the first part of (A1). Selecting $\pi(\xi)=\col(\xi,0)$ it is easy to see that the manifold invariance condition (A2) holds with the constant control
$$
c(\pi(\xi))=\lef[{c} - \theta \\ 0\rig].
$$
Moreover, $\pi(0)=\col(0,0)$, verifying the second part of (A1). The implicit manifold condition (A3) is verified with $\phi(x)=x_2$. Finally, we need to define a controller $v(x,z)$ for the augmented system
$$
\ba{rcl}
\dot x_1 &=& -x_1 +  \theta + v_1(x,z)\,\\
\dot x_2 &=& v_2(x,z)\\
\dot z &=& \nabla\phi(x) \lef[{c} -x_1 +  \theta + v_1(x,z) \\ v_2(x,z) \rig]=v_2(x,z),
\ea
 $$
that ensures boundedness of trajectories and $\lim_{t \to \infty} z(t)=0$. This is clearly guaranteed with the selection
$$
v(x,z)=\lef[{c} 0 \\ -z\rig].
$$

It is claimed in Theorem 2.1 of \cite{astkarort} that applying the control
$$
u=v(x,\phi(x))= \lef[{c} 0 \\ - x_2\rig],
$$
to the $x$--system ensures the origin is a GAS equilibrium. But the resulting closed--loop system
$$
\ba{rcl}
\dot x_1 &=& -x_1 +  \theta \\
\dot x_2 &=& - x_2
\ea
 $$
has a GAS equilibrium at $( \theta,0)$, not at the origin.

The source of the problem is that, if we do not impose  in (A4) the initial condition \eqref{zzer} we have only that $\dot z = \dot \phi=\dot x_2$ but $z(t) \neq \phi(x(t))=x_2(t)$. Indeed, integrating the system
$$
\ba{rcl}
\dot x_2 &=& -z \\
\dot z &=& -z,
\ea
 $$
 we get
\begequ
\lab{xzsys}
\ba{rcl}
z(t) \!\! &\!\!   =\!\!   &\!  e^{-t}z(0)\,\\
x_2(t)\!\! & \! \!=\! \! &\!  x_2(0)-\int_0^t e^{-\tau}z(0)d\tau=x_2(0)-z(0)(1-e^{-t})\,.
\ea
\endequ
Notice that the condition  \eqref{v} is also not verified since
$$
v(\pi(\xi),0)=\lef[{c} 0 \\ 0\rig] \neq c(\pi(\xi))=\lef[{c} - \theta \\ 0\rig].
$$

It is clear from \eqref{xzsys} that imposing the initial condition \eqref{zzer}, that is, $z(0)=x_2(0)$ we get $z(t)=x_2(t)$. But we still need to modify the controller to comply with   \eqref{v}. A simple choice being
$$
v(x,z)=\lef[{c} -\theta \\ -z\rig].
$$
For this new choice the control
$$
u=v(x,\phi(x))=\lef[{c} -\theta \\ -x_2\rig]
$$
yields $\dot x=-x$, which certainly has a GAS equilibrium at zero.

%
%%%%%%%%%%%
\subsection{Proof of Proposition \ref{pro2}}
\lab{appb}
%%%%%%%%%
%

The proof is divided in four parts establishing  global attractivity of (I) the manifold and (II) the equilibrium point, (III) local stability and (IV) hyperbolicity of the equilibrium point.
%
%%%%%
\subsection*{I. Global attractiveness of $\calM$}
Take $|\delta x|_x := \sqrt{V(x,\delta x)}$.  Given any differentiable curve $\gamma:[0,1] \to \rea^n$
define the horizontal length $\ell(\gamma) := \int_0^1 |\dot{\gamma}(s)|_{\gamma(s)} ds$.
Note that $\ell(\gamma) \neq 0$ iff $R(\gamma(s))\dot{\gamma}(s) \neq 0$ for some $s \in [0,1]$.
Thus,  $\ell(\gamma) \neq 0$ if $\gamma(s) \notin \calM$ for some $s\in [0,1]$.
For instance, consider any $x$ and $y$ in $\calM$. By construction,
there exists a differentiable curve $\gamma$ such that
$\gamma(0) = x$, $\gamma(1) = y$ and $\gamma(s) \in \calM$ for all $s\in [0,1]$.
Then $R(\gamma(s))\dot{\gamma}(s) = \nabla\phi(\gamma(s))\dot{\gamma}(s) = 0$,
by \eqref{eq:cond_R}, which implies $|\dot{\gamma}(s)|_{\gamma(s)}=0$, thus $\ell(\gamma) = 0$.
In a similar way, consider any $x \in \calM$ and $y \notin \calM$ and
let $\gamma$ be any differentiable curve such that
$\gamma(0) = x$ and $\gamma(1) = y$. Then, because of the rank condition on $R(x)$ and differentiability of $R(x)$,
there exists a measurable subset of
$\mathcal{I} \subset [0,1]$ such that $|\dot{\gamma}(s)|_{\gamma(s)} \neq 0$.
Thus, $\ell(\gamma) > 0$.

Let $\psi_t(x_0)$ denotes the flow of the system $\dot{x} = f(x) + g(x)\beta(x)$ at time $t$
from the initial condition $\psi_{0}(x_0)=x_0 \in \rea^n$.
Exploiting the boundedness of trajectories,
global attractiveness of $\mathcal{M}$ can be proven by showing that
$\lim_{t\to\infty} \ell(\psi_t(\gamma)) = 0$ for any given curve $\gamma$ such that $\ell(\gamma) \neq 0$.
We show this in the next two paragraphs.

By boundedness of trajectories, for any $\gamma:[0,1]\to \rea^n$ there exists a compact set $\mathcal{K}$ such that,
$\psi_t(\gamma(s)) \in \mathcal{K}$ for each $s\in[0,1]$ and $t\geq 0$. By continuity, for each $x\in \mathcal{K}$ and
$\delta x \in \rea^n$, \eqref{eq:dotV} guarantees that there exists $\lambda>0$ such that $\dot{V}(x,\delta x) < -\lambda V(x,\delta x) < 0$.
It follows that
$$
V\left(\psi_t(\gamma(s)), \dfrac{d}{ds}\psi_t(\gamma(s))\right) \leq \exp(-\lambda t) V\left(\gamma(s), \frac{d}{ds}\gamma(s)\right)\,,
$$
which implies that
$$
\left|\dfrac{d}{ds}\psi_t(\gamma(s))\right|_{\psi_t(\gamma(s))} \leq \exp\left(-\dfrac{\lambda t}{2} \right) \left|\dfrac{d}{ds} \gamma(s)\right|_{\gamma(x)}\,.
$$
Thus, $\ell(\psi_t(\gamma)) \leq \exp(-\frac{\lambda t}{2} ) \ell(\gamma)$.

Suppose now that $\gamma(0) \in \mathcal{M}$ and $\gamma(1) \notin \mathcal{M}$. By (A2),
$\psi_t(\gamma(0)) \in \mathcal{M}$ for all $t\geq 0$ (manifold invariance).
Thus, the combination of $\lim_{t\to\infty } \ell(\psi_t(\gamma)) =0$ with boundedness of trajectories guarantees that
$\psi_t(\gamma(1))$ converges asymptotically to $\mathcal{M}$.
%
%%%%%%%%
\subsection*{II. Global attractiveness of $x_*$}
%%%%%%%%
%
By (A2), (A4') and the boundedness of trajectories we have that any trajectory of the closed loop system converges to the manifold $\phi(x) = 0$. Moreover, by (A1) and (A2), the manifold is invariant and internally asymptotically stable, hence all trajectories of the closed loop system converge to the equilibrium $x_*$.\footnote{This steps coincide with the ones of Theorem 2.1 of \cite{astkarort}. The same remark applies to the derivations of part III.}
%
%%%%%%%%%%
\subsection*{III. Local stability of $x_*$}
%%%%%%%%%%
%
To conclude the proof we need to show that $x_*$ is Lyapunov stable. Note that any trajectory of the closed loop system is the image through the mapping $\pi(\cdot)$
of a trajectory of the target system $\xi$. Moreover, for any $\varepsilon_1$, there exists $\delta_1$ such
that $|\xi(0)| \leq \delta_1$ implies $|\xi(t)| \leq \varepsilon_1$. Hence, by regularity of $\pi$,
for any $\epsilon >0$ there exists $\delta>0$ such that
$|\pi(\xi(0))| \leq \delta$ implies $|\pi(\xi(t))| \leq \varepsilon$.

%
%%%%%%%
\subsection*{IV. Hyperbolicity of $x_*$}
%%%%%%%
%

Without loss of generality take $x_* = 0$ and define
$S := \nabla \alpha(0)$, $\Pi := \nabla \pi(0)$, and $A := \nabla[f(0)+g(0)c(0)]$.
For semplicity denote the matrix $R(0)$ by $R$
and $P(0)$ by $P$. The linearization of the closed loop system computed on the fixed point
reads $\tilde{x} = A \tilde{x}$. $\tilde{\xi} = S \tilde{\xi}$ denotes the linearization of the target
system at the fixed point $\xi_* = 0$.

Now, $S$ is Hurwitz by assumption. The span of the columns of $\Pi$ and $R^\top$ define two orthogonal subspaces of the state space.
To see this, note that $\Pi\tilde{\xi} \in T_{x_*}\mathcal{M}$ therefore
$R\Pi \tilde{\xi} = \nabla \phi(0) \Pi \tilde{\xi} = 0$. Since $\xi$ has dimension $p$,
it follows that $R\Pi = 0$. Clearly, $\Pi^\top R^\top = (R\Pi)^\top = 0$. It follows that
the state of the linearized closed loop system can be decomposed as
$$
\tilde{x} = \Pi \tilde\xi + R^\top e
$$
where $e$ is a vector in $\rea^{(n-p)}$. In particular, take
$$e := (RR^\top)^{-1} R(x-\Pi\tilde{\xi}) = (RR^\top)^{-1} R \tilde x \ .$$
The invertibility of $RR^\top$ follows from the rank condition on $R$.
The last identity follows from orthogonality, i.e. $R\Pi = 0$.

Take any trajectory $\tilde{x}(\cdot)$. Then, there exist positive
constants $c_1,c_2,c_3, c_4, c_5,c_6$ such that
$$
\begin{array}{rcl}
|\tilde x(t)|
& \leq & c_1 |\Pi \tilde\xi(t) + R^\top e(t)| \\
& \leq & c_1 |\Pi \tilde \xi(t) | + c_1 |R^\top e(t) | \\
& \leq & c_1 |\Pi\tilde \xi(t) | + c_1 |(RR^\top)^{-1} R \tilde x(t)| \\
& \leq & c_1 |\Pi\tilde \xi(t) | + c_2 \sqrt{V(0, \tilde x(t))} \\
& \leq & c_3 |\tilde \xi(t)| + c_4 \exp(-\frac{\lambda t}{2}) \sqrt{V(0, \tilde{x}(0))} \\
& \leq & c_3 \exp(\lambda_{\max}(S)t) |\tilde \xi(0)| + c_5 \exp(-\frac{\lambda t}{2}) |e(0)| \\
& \leq & c_6 \exp(\max\{\lambda_{\max}(S),-\frac{\lambda t}{2}\}t ) |\tilde x(0)| \\

\end{array}
$$
where $\lambda_{\max}(S)$ is the largest eigenvalue of $S$
and $\lambda$ is the local decay rate of $V$ (part 1 of the proof). Finally, exponential stability of the linearization
implies local exponential stability of $x_*$ for the closed-loop dynamics.

\end{document}